# A New Route for the Determination of Protein Structure in Physiological Environment

M. Altissimo[1*], M. Kiskinova[1], R. Mincigrucci[1], L. Vaccari[1], C. Guarnaccia[2] and C. Masciovecchio[1*]

[1]Elettra Sincrotrone Trieste, S. S. 14 km 163 - 34149 Basovizza, Trieste Italy

[2]International Centre for Genetic Engineering and Biotechnology, Padriciano 99 – 34149 Trieste, Italy

Corresponding authors: matteo.altissimo@elettra.eu, claudio.masciovecchio@elettra.eu

**Abstract**

Revealing the structure of complex biological macromolecules, such as proteins, is an essential step for understanding the chemical mechanisms that determine the diversity of their functions. Synchrotron based x-ray crystallography and cryo-electron microscopy have made major contributions in determining thousands of protein structures even from micro-sized crystals. They suffer from some limitations that have not been overcome, such as radiation damage, the natural inability to crystallize of a number of proteins and experimental conditions for structure determination that are incompatible with the physiological environment. Today the ultrashort and ultra-bright pulses of X-ray free-electron lasers (XFELs) have made attainable the dream to determine protein structure before radiation damage starts to destroy the samples. However, the signal-to-noise ratio remains a great challenge to obtain usable diffraction patterns from a single protein molecule. We describe here a new methodology that should overcome the signal and protein crystallization limits. Using a multidisciplinary approach, we propose to create a two dimensional protein array with defined orientation attached on a self-assembled-monolayer. We develop a literature-based, flexible toolbox capable of assembling different proteins on a functionalized surface while keeping them under physiological conditions during the experiment, using a water-confining graphene cover.

## 1. Introduction

The determination of the structure of proteins as well as other macromolecules has historically been prerogative of x-ray crystallography. One of the technique's requirements is the growth of high-quality crystals, which need to be sufficiently large to efficiently diffract x-rays while withstanding radiation damage. This method suffers from two noteworthy constraints, making structure determination extremely difficult or sometimes impossible. The first problem is that many bio-molecules hardly form sufficiently large, high quality crystals or do not crystalize at all. These restrictions are most severe for large protein complexes, such as membrane proteins, which participate in a plethora of fundamental biological processes as receptors, transporters or enzymes, and are therefore responsible of many cellular dysfunctions and potential targets for new therapeutic agents. The second limitation is the unavoidable x-ray radiation damage. Crystal size and radiation damage are inherently linked, since reducing the dose delivered to a single molecule requires large crystals amplifying the signal through Bragg diffraction. Therefore, synchrotron-based experiments are usually performed with cryo-cooled crystals in order to reduce the mass-transport rate due to radiation damage.

One of the breakthroughs using the ultrafast and coherent nature of intense X-ray pulses generated by free-electron laser (FEL) is single-shot coherent diffractive imaging performed before the destruction of the object, with diffraction-limited resolution. In life sciences, the ultimate goal for FELs is to solve the structure of large biomolecules without crystallization. This ambitious goal calls for the development of new experimental methodologies[1,2]. The first attempts to solve the structure of large biomolecules include the development of serial femtosecond crystallography (SFC) at room temperature, as well as the use of jets for continuous delivery of micro/nano crystals into the FEL hit zone. The structure is subsequently determined from the collection of many thousands diffraction patterns[3–5]. Time-resolved SFC experiments highlighted photo-induced changes in the electronic structure due to charge transfer, encoded in the diffraction pattern of the microcrystals under examination [6,7]. Interestingly, serial nano-crystallography has also become feasible with focused beams at synchrotrons[8], while microfluidic devices for protein crystallization and on-chip diffraction studies have been developed and demonstrated[9]. One of the main limitations for a broad applicability of SFC is that thousands of crystals are needed to obtain complete data sets. Considering that only a relatively small fraction of successful hits generates useful diffraction patterns, in order to become a routine analytical tool, SFC still lacks an easy, reliable and inexpensive way of producing a high number of nanocrystals.

The limited access to perfect crystals has been an unsolved problem in protein crystallography for decades, further adding to the difficulty in solving the structure of complex bio-molecules. It is interesting to note that some recent diffraction studies performed on XFEL on imperfect crystals have demonstrated that this is not a constraint. The 'neglected' weak continuous scattering arising when the crystals become disordered contains key information to overcome the resolution limits and to solve the tricky phase puzzle [10]. As shown in the paper, the continuously modulated background fully encodes the waves diffracted from individual single molecules. Thus, by using coherent diffraction imaging methods, it is possible to achieve real-space images of macromolecules with higher lateral resolution than what is obtainable by ordinary Bragg diffraction.

A common feature of all the achievements summarized above is that all the proteins structures solved at synchrotrons or FELs sources are based on the use of 3D crystals. It is interesting to note how, even in the case of 'in-vivo' room temperature SFC experiments, the protein crystal does not represent the protein in its more natural 2D arrangement. Until now 2D-crystallography has almost exclusively been the area of transmission electron microscopy (TEM). Significant progress in protein structure determination and lipid interaction has been made thanks to the use of cryo-TEM and the development of algorithms for recovering amplitude and phase information from recorded TEM images[11,12]. Given the reduced amount and fixed-target sample delivery in near-native environment, single-shot 2D protein crystallography with FELs was suggested as an attractive alternative[13,14] and has been recently demonstrated by proof-of-principle experiments at LCLS. One can also argue that the 2D approach adds the ability to explore protein function and dynamics and it is also an intermediate step towards the extremely ambitious XFELs case - atomic imaging of individual bio-molecules.

As in the 3D case, sample preparation is also a crucial step for the 2D case, where fixed-target solutions have to be developed. For 2D cryo-TEM studies, most often the proteins crystals are grown embedded within a lipid bilayer, whereas the first XFEL experiments used dices of silicon nitride windows for harvesting 2D crystals, which are then covered with a thin carbon film. These first experiments indicated that in order to overcome the present resolution limits of 7 Å and truly exploiting the unique XFEL properties, an improvement of both sample preparation and data analysis is crucial.

In the present work we put forward a new sample delivery method based on fabricating sample supports for hosting the target protein in a near-native environment. Our approach makes use of patterned silicon nitride membranes, over which a graphene cover ensures the stability of the liquid layer hosting the Protein Of Interest (POI). This is in turn covalently bound onto a chemically modified surface, so that an ordered array of proteins is produced in a hydrated environment and kept as close as possible to physiological conditions during data acquisition. The chemical binding method of the POI to the surface must ensure that the layer is 2D-ordered, and the protein in its biologically active state. Further, we present an experimental approach based on free electron laser diffraction before destroy, the final aim of our work being the determination of protein structure in physiological conditions.

## 2. Discussion

### a. Directed protein immobilization

The immobilization of a POI on a surface can be accomplished through several methods, both physical and chemical. Although physical methods are in general easier and more straightforward to develop, they tend to yield a disordered array of the POI on the substrate of choice. The added constraint of having and ordered array severely restricts the available choices. The binding process should satisfy the following criteria:

1) It covalently binds two unique and mutually reactive groups, one on the protein and one on the surface;
2) It is bio-orthogonal, i.e. it should not appear in, or cross-react with, any of the groups of endogenous amino acids;
3) It works efficiently under physiological conditions, and without the use of harsh chemicals that could cause the denaturation of the POI;
4) It does not alter substantially the structure of the POI, so as to leave it in its functional state;

The added complication is that the structure of each protein varies hugely; this impacts directly on the fact that there is no single method that can be used universally to attach proteins to a surface. Thus, given the POI, the binding methodology should be chosen between a set of several, i.e. a "toolbox". Since the final goal is to determine the structure of a protein, a corollary of points 1) and 4) is that, regardless of the chemistry used to bind the POI to a surface, the reacting moiety should be small in comparison to the POI. Engineering the positioning of proteins on a surface in ordered arrays has been studied extensively by the protein biochip community[15,16]. The POI will have to be modified with a unique chemical group or sequence at a site-specific location in order to ensure a covalent and oriented binding to the surface, which in turn must also be adapted to "receive" the POI. The adaptation of the POI can be accomplished through its recombinant expression by a suitably modified vector. Several classes of chemical reactions can be employed to engineer proteins onto a surface, and are reviewed in a number of papers[15–19].

The reactions listed in literature are:

a) catalyzed cyclo-addition [20,21], where the ring strain energy of a cyclooctine group catalyzes an

azide-alkyne cycloaddition, with the azide group attached to the POI.
b) modified Staudinger ligation [22,23] creating a stable amide bond between a POI-side azide group and an ester group attached to the surface.
c) Diels-Adler cyclo-addition[24,25], where a six-membered unsaturated ring is formed by a pair of dienophile/conjugate diene.
d) thiol-ene additions[26,27] where a thiol is added to an "-ene" group via a free radical mechanism, usually initiated by UV irradiation;
e) oxime ligation[28,29], which is the aniline-catalized condensation of an oxyamine or hydrazide with an aldehyde or a ketone, producing a stable oxime linkage.
f) Expressed Protein Ligation, such as Split-intein-mediated ligation[30,31] or farnesyltransferase-related methods[32]. While these are not strictly a single chemical reactions, they are used to bind a POI to a surface by means of the naturally-occurring protein trans-splicing process[33] (split-intein mediated case), or farnesyltransferase mediated binding[19].

All these reactions satisfy the 4 requirements listed above, and therefore they constitute an appropriate "toolbox" for anchoring ordered 2-D arrays of proteins on a substrate. Their effectiveness has already been reported[34].

Among the methods described above, we will concentrate on the split-intein mediated ligation, since it is able to yield a traceless attachment of the POI to a gold surface, with minimal POI modification. The technique has been first described in[30]. The basic idea described there is to genetically engineer a suitable expression vector, with the most common hosts being *E. Coli, P. Pastoris* or *S. Cerevisiae,* to express the POI with an N-intein fragment attached to its C-terminal. On the surface, the complementary C-intein fragment is attached to a modified polyethylene glycol (PEG) linker. When the C-intein and the N-intein fragments interact, they form an active intein domain, binding the POI to the surface. The split intein is naturally spliced into the solution, leaving the POI attached to the surface. Generally speaking, a Self-Assembled Monolayer (SAM) of two differently modified PolyEthylene Glycol (PEG) chains are used, given that they are readily available and chemically well understood[35]. The shorter of the two is used as a spacer, separating the substrate from the solution, while the second longer one is modified to bind with the POI.

b. **Proposed methodology**

The tight integration of sample delivery system and of POI orientation will maximize the chance of a successful protein structure reconstruction. We intend to use the toolbox on different kind of protein in order to demonstrate the potential of our approach. At first, we focus on Human Serum Albumin (HSA), in order to show how the proposed method would allow investigating the structure of a well-characterized system in physiological environment, a condition that is not compatible with standard protein diffraction methods. HSA can been expressed in high yield in the culture medium by the methylotrophic yeast Pichia pastoris both alone or as a fusion protein (see [36–38]). The HSA structure has been resolved in 1999 by Sugio and coworkers[39]; all the relevant protein data can be found at the Protein Data Bank[40]. HSA has an exposed C-terminal that allows for a N-intein group to be attached through recombinant techniques, see Figure 1.

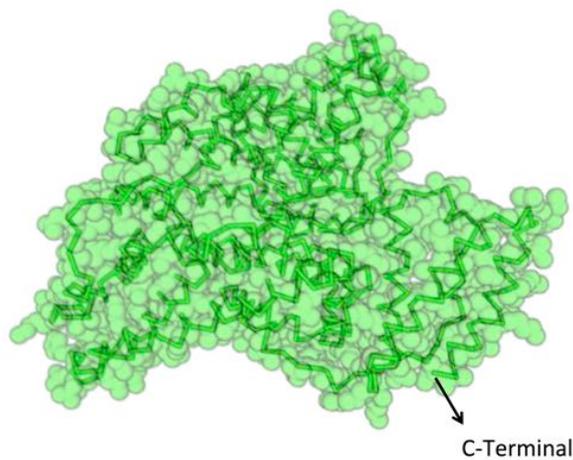

Figure 1: Human Serum Albumine. The backbone of the protein is shown superimposed to a space fill model. The C-terminal of the protein is also indicated. Data is from [39,40].

Further, being a protein found in blood, it is safe to assume that the protein will be in its functional conformation in a physiological solution. The protein will be bound on Au, through a tailored SAM. The 10 nm Au layer is going to be deposited on a silicon nitride membrane, and the underlying silicon chip will host a number of spatially separated membranes, so as to line each one up with a single FEL shot and collect the corresponding diffraction pattern. Moving the Si chip to the next window, and repeating the process on another membrane will allow to collect the diffraction pattern in a step-and-repeat fashion. A graphene cover will prevent the evaporation of the physiological solution in the vacuum chamber[41]. This would be ideal, due to the transparency of the graphene layer to FEL photons. As an alternative, a second set of nitride membranes could be used for sealing the wet chamber.

Standard photolithography techniques will be used to open silicon nitride membranes and define 10 nm thick, 10 µm in side Au square patches, using 2 nm Ti as an adhesion layer. The periodicity of the FEL-radiation transparent areas on the substrate is determined by the thickness of the Si substrate, due to the angle between the Si (111) and (100) planes. In order to provide the graphene cover with a set of supports, an optical resist will be patterned via photolithography to "corral" each of the membranes. The photoresist props also serve to keep the POI environment wet and to create separated micro-environments, in order to minimize the possibility of water leakage after each FEL shot. Figure 2 schematizes the end result, excluding the graphene cover.

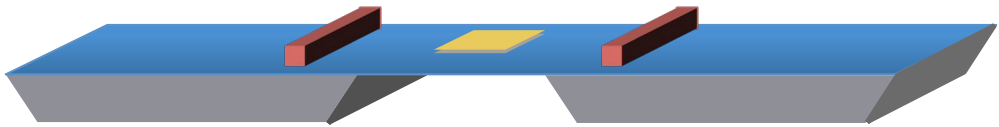

Figure 2: schematics of a single FEL-transparent area. The silicon nitride membrane (in light blue) is used as a carrier for the 10 micron wide, 10 nm thick Au square (in gold). Outside of the membrane perimeter, the area is corralled by patterned photoresist, used as a prop for the graphene cover (not shown in the figure) and to keep each sample area wet, and separated from the others.

To define the SAM, we will proceed as described in [19] with a slight modification. A dithiocarbamate (DTC)-modified PEG will serve as a linker, with the loose end (i.e. the one not attached to the underlying gold layer) modified by a C-intein domain as described in [16,30]. A shorter, unmodified DTC-PEG layer will be used as a spacer. The density of the anchoring points can be controlled by adjusting the relative concentration of the two DTC's in solution. The POI (specifically HSA) will be recombinantly over-expressed by genetically engineered Pichia Pastoris[42] or by other suitable eukaryotic vector. One of the key points of the binding process proposed here is that the POI will bind to the surface directly from the cell culture medium thus avoiding lengthy purification steps. The POI

will be covalently linked to the Au surface via the intein-mediated trans-splicing process as described

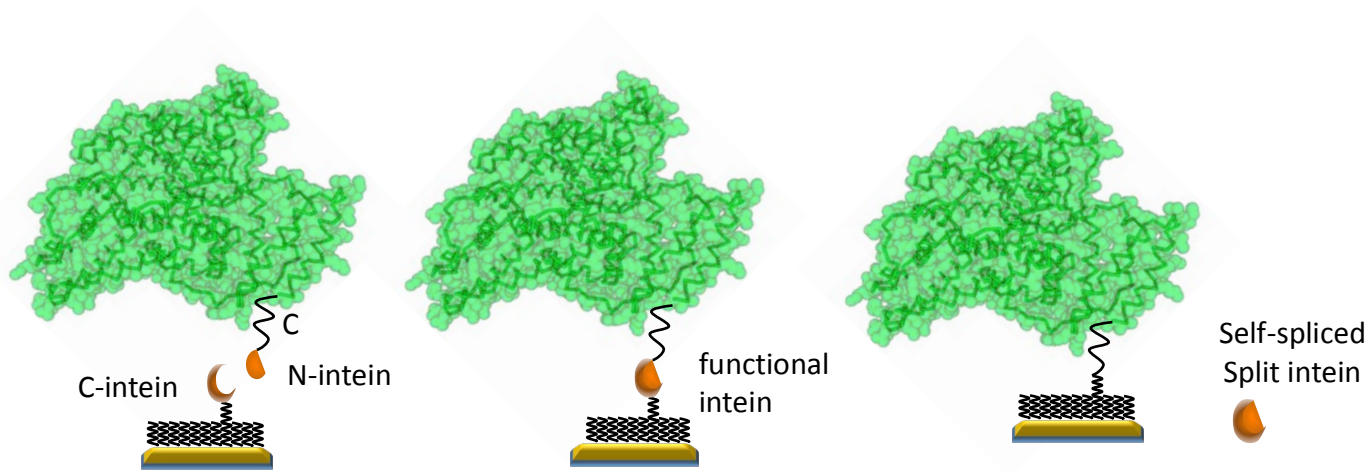

above.

Figure 3: schematics of HSA covalent binding process to the surface.

All the other cell debris will be washed away with an appropriate rinsing step. Figure 3 shows schematically the binding process. By rough calculation, according to the data published on the Protein Data Bank, we estimate that there will be approximately 2.5 million HSA proteins per 10 μm square Au patch. Once the proteins have been attached to the surface, a graphene layer will be added on top to seal the stack. What we believe is particularly interesting about our methodology is the fact that any protein of known sequence can potentially be recombinantly modified to incorporate into either the N- or the C-terminal a suitable moiety that will bind to a chemically complementary substrate, regardless of the ability of the POI to crystallize. As stated above, the density of the surface anchoring points will be determined by the relative concentration of the spacer/linker PEG chains. Coupled to this, the protein-protein interactions due to charge distribution and steric hindrance will order of the POI onto the chemically-engineered surface. This has been demonstrated in literature by, for example[19,43].

## c. Experimental Needs

In this section we will discuss the experimental requirements needed to reconstruct the protein structure. To avoid sample degradation due to radiation damage, FELs pulses represent a valuable choice[44]. Accordingly to[45], diffraction images can be easily acquired to better than 8.5 Å resolution, from two-dimensional protein crystal samples less than 10 nm thick and at room temperature. While this achievement represents a big step in the direction of determining protein structure in physiological condition, the limits of the approach proposed in [45] were basically related to the fact that the protein had random orientation on the substrate. As the authors pointed out: "The extent to which multiple lattice or powder diffraction XFEL data can be reliably used for structure determination from 2-D crystals is currently unclear". Using our approach in the case of HSA we can align up to about $2.5 \cdot 10^8$ proteins on a silicon nitride substrate as large as 100 mm².

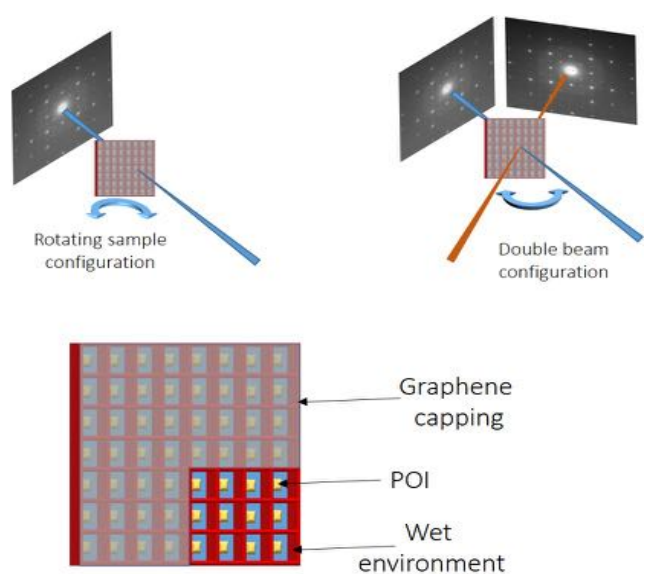

Figure 4 – Proposed set-up for the single beam diffraction (top left) and for double beam diffraction (top right). The angle of incidence of the beam with respect to the surface and/or the inter-beam angle can be varied to obtain complementary diffraction patterns.

This will strongly increase the intensity from Bragg diffraction and, moreover, the use of a large focal spot will prevent electron stripping from the atoms upon pulse arrival, thus decreasing the scattering from unbound electrons and further increasing the signal-to-noise ratio. Finally, the use of a graphene cover allows keeping the protein in physiological condition. Focal spot sizes can be easily varied from 10 μm to 1 mm using standard optics set-ups, thus defining the maximum POI island size that can be illuminated by a single shot. In case of ablation, samples can be raster scanned. The diffraction pattern of each POI patch may be acquired with a suitable in-vacuum area detector. When single shot techniques have been applied to nano-crystals, the 3D structure of the protein has been recovered by sorting and merging diffraction patterns obtained by randomly oriented crystals in liquid jets[3]. With our approach, one can rotate the sample under the beam so as to precisely define the angle under which the POI is observed. Alternatively, a set-up consisting of two crossed FEL beams[46] (Figure 4) and two detectors may be employed to simultaneously record the diffraction pattern of the same area from two different directions.

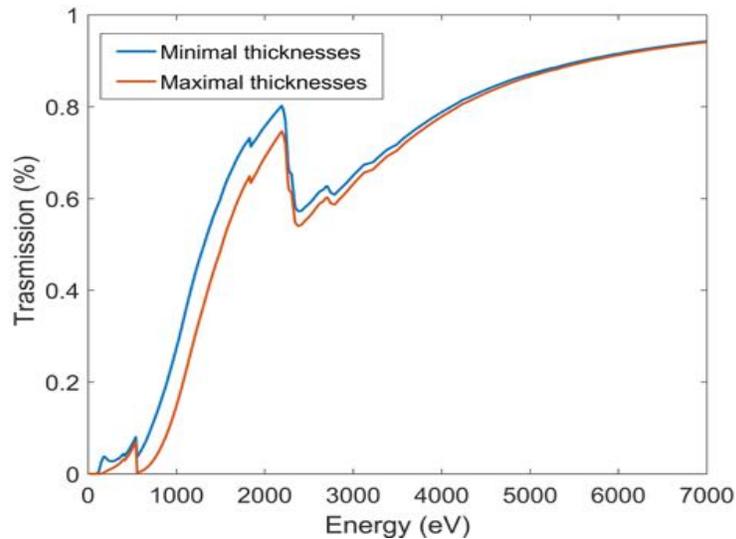

Figure 5- Estimated sample transmission for 50 nm of Silicon Nitride and 0.5 mm of water (blue curve) and for 100 nm of Silicon Nitride and 2 mm of water (orange curve).

Figure 5 shows the calculated theoretical transmission of the sample stack. The blue and the orange curve refer to 50 nm nitride/0.5 μm water, and 100 nm nitride/1 μm water respectively. For both of the curves, a 2 nm Ti and 10 nm Au bilayer has been considered. As expected, a monolayer of proteins does not contribute substantially to the total absorption of the stack. The proposed structure has a static transmission value around 40 % in the least favorable case for photon energies above 1 keV.

In order to calculate the signal improvement caused by the orientation of the POI, it is interesting to estimate the increase in the Bragg signal due to the number of proteins aligned on the surface and under the beam. As in any interference problem, the strength of the main peak is proportional to the number of diffracting objects squared[47], while the ratio between different order/directions is governed by more complex relations[48]. For photosystem I as studied by[3] with single FEL shot methods, an intensity in the Bragg spots ranging from $5 \times 10^3$ to 5 photons was recorded as function of angle. The lowest reported crystal size (P6$_3$ crystal structure with a diameter of 160 nm and a diameter to length ratio of 1:2) corresponds to a number of proteins ranging from 270 to 540. With our sample delivery method, used for the same photosystem I, we expect to obtain a Bragg peak intensity which is at least a factor 6 x 10$^6$ higher, since a POI SAM of 100 x 100 mm should contain ~ 1.3 x 10$^7$ proteins. Similar ratios are thus expected for others proteins.

The recent development of FEL Facilities has opened up different schemes for the production of photon pulses[49]. One recently highlighted possibility is to allow only a single superradiant spike of the electron bunch self-emission to be amplified in an exotic configuration. This mode may produce pulses as short as 500 attosecond, with the drawback of a limited number of photons per pulse. Using the parameters of the European XFEL under commissioning in Hamburg[50], the expected flux would be about 10$^{11}$ photon per pulse. It has been demonstrated that FEL single molecule imaging will strongly benefit from using sub-femtosecond pulses, due to the significant radiation damage and the formation of preferred multisoliton clusters, which reshape the overall electronic density of the molecular system when using longer pulses[51]. While the expected number of photons per pulse in the superradiance scheme would be too low to get the diffraction pattern from a single molecule, this is not the case for the 2D array sample delivery method we introduce here, thanks to the several orders of magnitude enhancement of the scattering power due to the large number of molecules involved in the process.

3. **Conclusions**

In this article we propose a new method capable of delivering two-dimensional, aligned arrays of bio-molecules. The alignment should increase the signal-to-noise ratio by several orders of magnitude on FEL-based diffraction experiments, as compared with other reported studies. The sample will be a two-dimensional crystal containing up to billions of ordered macromolecules that will be kept in a near-native environment. The route presented above follows a multidisciplinary approach based on existing literature. We discussed in detail the advantages of this methodology that we highlight in the following:
1) It will permit to reveal the structure of macromolecules in general, and proteins more in particular, that cannot be crystallized and, therefore, cannot be studied by classical crystallography;
2) The 2D protein array would allow to use FEL pulses produced in exotic configurations with shorter pulse duration (namely few attosecond). This would definitively avoid sample radiation damage during the diffraction process.

3) The measurements can be carried out under physiological conditions;
4) Only a moderate sample quantity will be necessary. This is extremely important when samples are difficult or very expensive to produce.

Moreover, our approach avoids lengthy purification steps of the POI, due to the bio-orthogonal chemical reaction binding it to the engineered surface.

It is worthwhile noting that a different ordering strategy is required in order to solve the structure of membrane proteins in their native environment. These constitute 20-30% of the total proteome, but represent about 1% of determined protein structures, due to difficulties in their crystallization. Cryo-TEM has been one of the main tools of choice for integral membrane proteins (IMP) structure reconstruction. The base of the 2D crystallization behind cryo-TEM IMP structure reconstruction is a work recognizing as early as 1992[52] that IMP can form 2D crystals when inserted into an artificial phospholipid bilayer [53,54]. Due to the nature of IMPs, ad-hoc crystallization conditions have to be determined for each protein[55–57]. A toolbox for the 2D crystallization of IMPs is therefore already well established, and we suggest that FEL-based in particular, and more in general x-ray, IMP structure reconstruction should follow the methodologies found in literature.

An engineered sample delivery method can be used to exploit the two main characteristics of FEL pulses, i.e. tunability and coherence. The former can be used to perform anomalous scattering experiments across the edge of biologically abundant elements, such as S and P. The coherence of the beam, coupled with our sample holder, will allow for precisely reconstructing the phase of the scattered photons.

We note finally that the sample delivery method we propose here allows for pump-and-probe experiments, whereby a visible laser excites the POI, and the FEL beam probes its structure post-excitation. This could, for example, enable to determine the dynamic change in the structure of a photosynthetic centre, before, during and after its excitation by visible-light.

## 4. Acknowledgments

The authors acknowledge G. Kourousias for useful discussions on phase retrieval methods.

# 5. References

1. Neutze, R., Brändén, G. & Schertler, G. F. X. Membrane protein structural biology using X-ray free electron lasers. *Curr. Opin. Struct. Biol.* **33,** 115–125 (2015).

2. Kern, J., Yachandra, V. K. & Yano, J. Metalloprotein structures at ambient conditions and in real-time: Biological crystallography and spectroscopy using X-ray free electron lasers. *Curr. Opin. Struct. Biol.* **34,** 87–98 (2015).

3. Chapman, H. N. *et al.* Femtosecond X-ray protein nanocrystallography. *Nature* **470,** 73–77 (2011).

4. Gallat, F.-X. *et al.* In vivo crystallography at X-ray free-electron lasers: the next generation of structural biology? *Philos. Trans. R. Soc. B Biol. Sci.* **369,** 20130497–20130497 (2014).

5. Schlichting, I. Serial femtosecond crystallography: The first five years. *IUCrJ* **2,** 246–255 (2015).

6. Kern, J. *et al.* NIH Public Access. **340,** 491–495 (2013).

7. Kupitz, C. *et al.* Serial time-resolved crystallography of photosystem II using a femtosecond X-ray laser. *Nature* **513,** 261–265 (2014).

8. Gati, C. *et al.* Serial crystallography on in vivo grown microcrystals using synchrotron radiation. *IUCrJ* **1,** 87–94 (2014).

9. Heymann, M. *et al.* Room-temperature serial crystallography using a kinetically optimized microfluidic device for protein crystallization and on-chip X-ray diffraction. *IUCrJ* **1,** 349–360 (2014).

10. Ayyer, K. *et al.* Macromolecular diffractive imaging using imperfect crystals. *Nature* **530,** 202–206 (2016).

11. Hite, R. K., Raunser, S. & Walz, T. Revival of electron crystallography. *Curr. Opin. Struct. Biol.* **17,** 389–395 (2007).

12. Gipson, B. R. *et al.* Automatic recovery of missing amplitudes and phases in tilt-limited electron crystallography of two-dimensional crystals. *Phys. Rev. E - Stat. Nonlinear, Soft Matter Phys.* **84,** 1–9 (2011).

13. Kewish, C. M., Thibault, P., Bunk, O. & Pfeiffer, F. The potential for two-dimensional

crystallography of membrane proteins at future x-ray free-electron laser sources. *New J. Phys.* **12,** (2010).

14. Proteomics, C. *Architecture*.

15. Steen Redeker, E. *et al.* Protein engineering for directed immobilization. *Bioconjug. Chem.* **24,** 1761–1777 (2013).

16. Camarero, J. A. Recent developments in the site-specific immobilization of proteins onto solid supports. *Biopolym. - Pept. Sci. Sect.* **90,** 450–458 (2008).

17. Rusmini, F., Zhong, Z. & Feijen, J. Protein immobilization strategies for protein biochips. *Biomacromolecules* **8,** 1775–1789 (2007).

18. Chen, Y. X., Triola, G. & Waldmann, H. Bioorthogonal chemistry for site-specific labeling and surface immobilization of proteins. *Acc. Chem. Res.* **44,** 762–773 (2011).

19. Choi, S. R., Seo, J. S., Bohaty, R. F. H. & Poulter, C. D. Regio- and chemoselective immobilization of proteins on gold surfaces. *Bioconjug. Chem.* **25,** 269–275 (2014).

20. Baskin, J. M. *et al.* Copper-free click chemistry for dynamic in vivo imaging. *Proc. Natl. Acad. Sci. U. S. A.* **104,** 16793–7 (2007).

21. Agard, N. J., Prescher, J. A. & Bertozzi, C. R. A strain-promoted [3 + 2] azide-alkyne cycloaddition for covalent modification of biomolecules in living systems. *J. Am. Chem. Soc.* **126,** 15046–15047 (2004).

22. Staudinger, H. & Hauser, E. ??ber neue organische Phosphorverbindungen IV Phosphinimine. *Helv. Chim. Acta* **4,** 861–886 (1921).

23. Saxon, E. & Bertozzi, C. R. Cell Surface Engineering by a Modified Staudinger Reaction. *Science (80-. ).* **287,** 2007–2010 (2000).

24. Houseman, B. T., Huh, J. H., Kron, S. J. & Mrksich, M. Peptide chips for the quantitative evaluation of protein kinase activity. *Nat. Biotechnol.* **20,** 270–274 (2002).

25. De Araújo, A. D. *et al.* Diels-Alder ligation and surface immobilization of proteins. *Angew. Chemie - Int. Ed.* **45,** 296–301 (2005).

26. Poster, T. No Title. *Ber. Deutsch. Chem. Ges.* **38,** 646−657 (1905).

27. Hoyle, C. E. & Bowman, C. N. Thiol-ene click chemistry. *Angew. Chemie - Int. Ed.* **49,** 1540–1573 (2010).

28. Kaumaya, P. *Protein engineering*. (InTech, 2012).

29. Dirksen, A. & Dawson, P. E. Rapid oxime and hydrazone ligations with aromatic aldehydes for biomolecular labeling. *Bioconjug. Chem.* **19,** 2543–2548 (2008).

30. Kwon, Y., Coleman, M. A. & Camarero, J. A. Selective immobilization of proteins onto solid supports through split-intein-mediated protein trans-splicing. *Angew. Chemie - Int. Ed.* **45,** 1726–1729 (2006).

31. Braner, M., Kollmannsperger, A., Wieneke, R. & Tampé, R. 'Traceless' tracing of proteins – high-affinity trans-splicing directed by a minimal interaction pair. *Chem. Sci.* 2646–2652 (2016). doi:10.1039/C5SC02936H

32. Gauchet, C., Labadie, G. R. & Poulter, C. D. Regio- and chemoselective covalent immobilization of proteins through unnatural amino acids. *J. Am. Chem. Soc.* **128,** 9274–9275 (2006).

33. Wu, H., Hu, Z. & Liu, X. Q. Protein trans-splicing by a split intein encoded in a split DnaE gene of Synechocystis sp. PCC6803. *Proc Natl Acad Sci U S A* **95,** 9226–9231 (1998).

34. Lin, P. C., Weinrich, D. & Waldmann, H. Protein biochips: Oriented surface immobilization of proteins. *Macromol. Chem. Phys.* **211,** 136–144 (2010).

35. Samanta, D. & Sarkar, A. Immobilization of bio-macromolecules on self-assembled monolayers: methods and sensor applications. *Chem. Soc. Rev.* **40,** 2567–2592 (2011).

36. Zhu, R.-Y. *et al.* Expression and purification of recombinant human serum albumin fusion protein with VEGF165b in Pichia pastoris. *Protein Expr. Purif.* **85,** 32–7 (2012).

37. Kobayashi, K. *et al.* High-level expression of recombinant human serum albumin from the methylotrophic yeast Pichia pastoris with minimal protease production and activation. *J. Biosci. Bioeng.* **89,** 55–61 (2000).

38. Belew, M., Yan, M. & Caldwell, K. Purification of Recombinant Human Serum Albumin (rHSA) Produced by Genetically Modified Pichia Pastoris. *Sep. Sci. Technol.* **43,** 3134–3153 (2008).

39. Sugio, S., Kashima, a, Mochizuki, S., Noda, M. & Kobayashi, K. Crystal structure of human serum albumin at 2.5 A resolution. *Protein Eng.* **12,** 439–446 (1999).

40. Protein Data Bank. Available at:

http://www.rcsb.org/pdb/explore/explore.do?structureId=1AO6.

41. Berry, V. Impermeability of graphene and its applications. *Carbon N. Y.* **62,** 1–10 (2013).

42. Chuang, V. T. G. and Otagiri, M. Recombinant human serum albumin. *Drugs of Today* **8,** 547 (2007).

43. Weinrich, D. *et al.* Oriented immobilization of farnesylated proteins by the thiol-ene reaction. *Angew. Chemie - Int. Ed.* **49,** 1252–1257 (2010).

44. Neutze, R., Wouts, R., van der Spoel, D., Weckert, E. & Hajdu, J. Potential for biomolecular imaging with femtosecond X-ray pulses. *Nature* **406,** 752–757 (2000).

45. Frank, M. *et al.* Femtosecond X-ray diffraction from two-dimensional protein crystals. *IUCrJ* **1,** 95–100 (2014).

46. Bencivenga, F. *et al.* Four-wave mixing experiments with extreme ultraviolet transient gratings. *Nature* **520,** 205–8 (2015).

47. Jackson, J. D. *Classical electrodynamics*. (1962).

48. Ashcroft, N. W. & Mermin, David, N. *Solid State Physics*. (1976).

49. Bonifacio, R., Fares, H., Ferrario, M., W. J. McNeil, B. & R. M. Robb, G. Design of sub-Angstrom compact free-electron laser source. *Opt. Commun.* **382,** 58–63 (2017).

50. The Technical Design Report (TDR) of the European XFEL. Available at: https://xfel.desy.de/technical_information/tdr/tdr/.

51. Fratalocchi, A. & Ruocco, G. Single-molecule imaging with X-ray free-Electron Lasers: Dream or Reality? *Phys. Rev. Lett.* **106,** 1–4 (2011).

52. Kühlbrandt, W. Two-dimensional crystallization of membrane proteins. *Q. Rev. Biophys.* **25,** 1–49 (1992).

53. Coutable, A. *et al.* Preparation of tethered-lipid bilayers on gold surfaces for the incorporation of integral membrane proteins synthesized by cell-free expression. *Langmuir* **30,** 3132–3141 (2014).

54. Giess, F., Friedrich, M. G., Heberle, J., Naumann, R. L. & Knoll, W. The Protein-Tethered Lipid Bilayer: A Novel Mimic of the Biological Membrane. *Biophys. J.* **87,** 3213–20 (2004).

55. De Zorzi, R. *et al.* Growth of large and highly ordered 2D crystals of a K+ channel, structural

role of lipidic environment. *Biophys. J.* **105,** 398–408 (2013).

56. Signorell, G. A., Kaufmann, T. C., Kukulski, W., Engel, A. & Rémigy, H. W. Controlled 2D crystallization of membrane proteins using methyl-??-cyclodextrin. *J. Struct. Biol.* **157,** 321–328 (2007).

57. Ubarretxena-Belandia, Iban, and Stokes, D. L. *Recent Advances in Electron Cryomicroscopy*. *Recent Advances in Electron Cryomicroscopy, Part A* **81,** (2010).